\documentclass[reprint, superscriptaddress, secnumarabic, amssymb, nobibnotes, showpacs, aps, prb]{revtex4-1}

\newcommand{\musr}{$\mu$SR}
\newcommand{\tc}{$T_\mathrm{c}$}

\newcommand{\ucf}{$B_\mathrm{c2}$}

\usepackage{amsmath}
\usepackage{graphicx}
\usepackage{epstopdf}
\usepackage[]{natbib}
\usepackage[separate-uncertainty=true]{siunitx}
\usepackage{multirow}

\begin{document}

\graphicspath{{./Figs/}}

\title{Probing the superconducting ground state of the rare-earth ternary boride superconductors $R$RuB$_2$ ($R$ = Lu,Y) using muon-spin rotation and relaxation}

\author{J.~A.~T.~Barker}
\email[]{Joel.Barker@warwick.ac.uk}
\affiliation{Physics Department, University of Warwick, Coventry, CV4 7AL, United Kingdom}
\affiliation{Laboratory for Muon-Spin Spectroscopy, Paul Scherrer Institut, CH-5232 Villigen PSI, Switzerland}

\author{R.~P.~Singh}
\affiliation{Department of Physics, Indian Institute of Science Education and Research Bhopal, Bhopal-462066, India}

\author{A.~D.~Hillier}
\affiliation{ISIS facility, STFC Rutherford Appleton Laboratory, Harwell Science and Innovation Campus, Oxfordshire, OX11 0QX, United Kingdom}

\author{D.~M$^\textrm{c}$K.~Paul}
\affiliation{Physics Department, University of Warwick, Coventry, CV4 7AL, United Kingdom}

\date{\today}

\begin{abstract}
The superconductivity in the rare-earth transition metal ternary borides $R$RuB$_2$ (where $R$ = Lu and Y) has been investigated using muon-spin rotation and relaxation.  Measurements made in zero-field suggest that time-reversal symmetry is preserved upon entering the superconducting state in both materials; a small difference in depolarization is observed above and below the superconducting transition in both compounds, however this has been attributed to quasistatic magnetic fluctuations.  Transverse-field measurements of the flux-line lattice indicate that the superconductivity in both materials is fully gapped, with a conventional \emph{s}-wave pairing symmetry and BCS-like magnitudes for the zero-temperature gap energies.  The electronic properties of the charge carriers in the superconducting state have been calculated, with effective masses $m^*/ m_\mathrm{e} = $ \num{9.8\pm0.1} and \num{15.0\pm0.1} in the Lu and Y compounds, respectively, with superconducting carrier densities $n_\mathrm{s} = $ (\num{2.73\pm0.04})~$\times 10^{28}$~\si{\per\meter\cubed} and (\num{2.17\pm0.02})~$\times 10^{28}$~\si{\per\meter\cubed}.  The materials have been classified according to the Uemura scheme for superconductivity, with values for $T_\mathrm{c}/T_\mathrm{F}$ of $1/(414\pm6)$ and $1/(304\pm3)$, implying that the superconductivity may not be entirely conventional in nature.
\end{abstract}

\pacs{74.25.Uv,74.25.Ha,74.70.Dd,76.75.+i}

\maketitle

\section{Introduction}
Rare-earth ternary boride superconductors are a class of materials which have been observed to exhibit relatively large values of the superconducting transition temperature, \tc.~\cite{Wolowiec2015}  The transition-metal borides with atomic formular $RT_4$B$_4$ (where $R$ is the rare-earth atom and $T$ is a transition metal) can crystallize in a number of polytypes, including primitive tetragonal,~\cite{Matthias1977} body-centred tetragonal,~\cite{Johnston1977} or orthorhombic crystal structures.~\cite{Yvon1982}  In all these polytypes, the boron atoms are found to have dimerized into non-interacting $B_2$ units.  The highest values of \tc\ have been found in the tetragonal polytypes, where the transition metal atoms cluster into isolated tetrahedra and form linear or zigzag chains.  In the orthorhombic structure, the $T$ atoms form an extended three-dimensional cluster that interpenetrates.  The superconducting transition temperatures are systematically lower in the orthorhombic polytype than the tetragonal compounds across the whole range of rare-earth elements, implying that the dimensionality of the $T$ clusters plays an important role in the superconductivity.~\cite{Johnston1982}

\begin{figure}
\centering
\includegraphics[width=1.0\columnwidth]{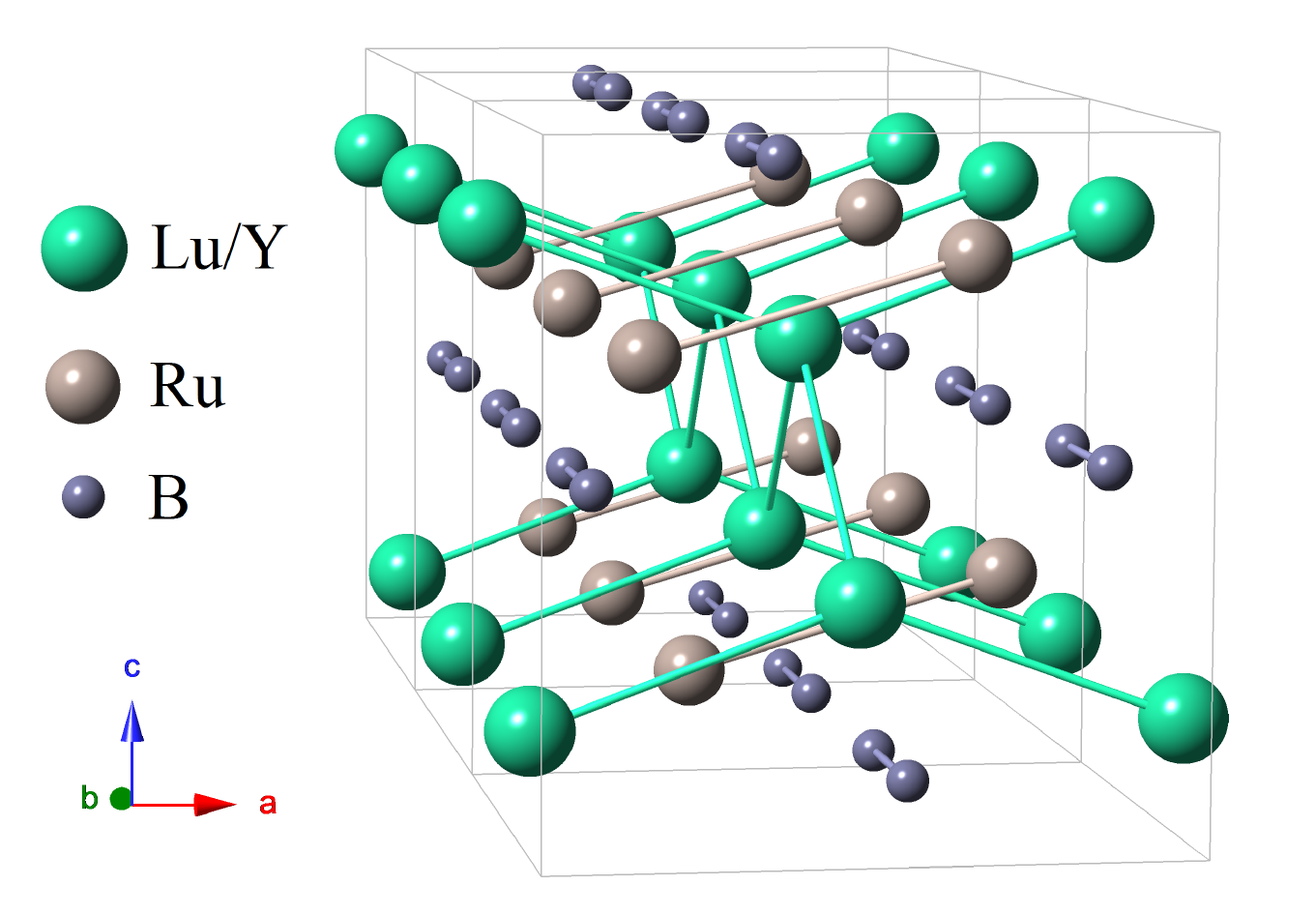}
\caption{(Color online) Crystal structure of the $R$RuB$_2$ ternary borides.  The $R$ atoms (large spheres) form zig-zag chains that run parallel to the $b$ crystallographic axis.  The B atoms (small spheres) form weakly interacting dimers, with the Ru atoms (medium spheres) isolated.}
\label{fig:structure}
\end{figure}

A new structural phase in the transition metal ternary boride family was reported in 1980, after anomalous superconducting transitions were observed with \tc's that did not match known structures.~\cite{Shelton1980}  The stoichiometrically distinct $RT$B$_2$ phase crystallizes into an orthorhombic structure with the $Pnma$ space-group.  The key feature in this material are zig-zag chains of rare-earth atoms, with dimerized boron.  The boron dimers weakly interact, forming straight chains that run in parallel to direction of the main $R - R$ zig-zag chain, and are perpendicular to planes of $R$ and $T$ atoms (see Fig.~\ref{fig:structure}).  Only compounds with non-magnetic $R$ atoms exhibit superonductivity, whereas the inclusion of magnetic atoms is accompanied by magnetic ordering with critical temperatures up to \SI{46}{\kelvin}.~\cite{Ku1980}

Two materials in this family, LuRuB$_2$ and YRuB$_2$, are important as reference materials for studying the entire family tree - the $4f$ electron shell is full in the Lu compound, and empty in the Y compound.  Superconductivity has been reported in the Lu compound at temperatures of \SI{9.7}{\kelvin} - \SI{10.1}{\kelvin}, and in the Y compound at temperatures of \SI{7.2}{\kelvin} - \SI{7.8}{\kelvin}, with large values for the upper critical field \ucf\ of \SI{5.7}{\tesla} and \SI{4.8}{\tesla}, respectively.~\cite{Ku1980,Lee1987}  These large values indicate that the superconductivity might be expected to be strongly coupled, with a high superconducting carrier density.  However, NMR measurements have identified that these materials appear to lie in the weak-coupling limit of the conventional BCS theory.~\cite{Kishimoto2009,Bardeen1957}  In this paper, we report the results of a muon-spin rotation and relaxation (\musr) study of the superconducting properties in this pair of materials.  We combine the results with previously reported findings in order to further characterize the electronic properties of the superconducting state.

The \musr\ technique provides an excellent means of characterizing superconductors, as it probes the magnetism in a sample at a microscopic level.  Spin-relaxation experiments in zero-field (ZF) allow the detection of spontaneous magnetization that can be associated with spin-triplet superconductivity.~\cite{Luke1998,Aoki2003,Hillier2009,Hillier2012,Bhattacharyya2015}  Because \musr\ measures the field distribution across the sample, the temperature dependence and absolute value of the magnetic penetration depth can be established to a high degree of accuracy.  Using this information, multiband superconductivity, line or point nodes, as well as anisotropy in the order parameter can all be unambiguously determined.~\cite{Biswas2011a,Sonier1994,Khasanov2008}  A key strength of \musr\ is that even in polycrystalline samples, the angular average is often enough to reliably observe these effects.

\section{Experimental Details}

\subsection{Sample preparation}

Polycrystalline samples of LuRuB$_2$ and YRuB$_2$ were prepared by arc-melting stoichiometric quantities of high-purity Y/Lu, Ru, and B in a tri-arc furnace under an Ar (5N) atmosphere on a water cooled copper hearth.  Each sample was flipped and remelted several times in order to improve the homogeneity of the as-cast ingot.  The samples were subsequently sealed in evacuated quartz tubes, and annealed at \SI{1050}{\celsius} for one week.  

\subsection{Sample Characterization}

\begin{figure}
\centering
\includegraphics[width=1.0\columnwidth]{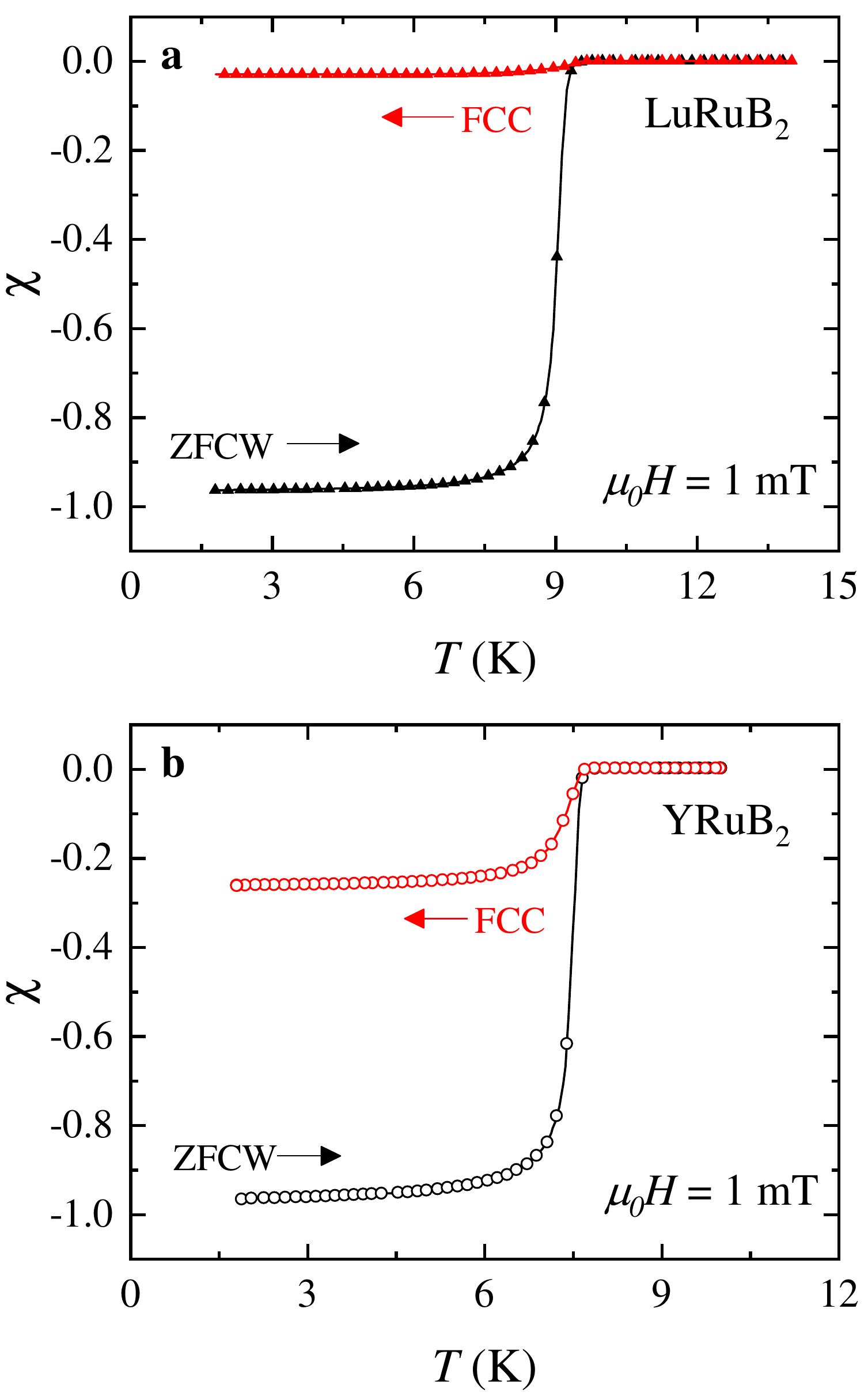}
\caption{(Color online) Temperature dependence of the magnetic susceptibility $\chi$ for (a) LuRuB$_2$ and (b) YRuB$_2$.  The samples were cooled in zero-field to \SI{1.8}{\kelvin}, at which point a field of \SI{1}{\milli\tesla} was applied.  Data were collected upon zero-field-cooled warming (ZFCW) and during a subsequent field-cooled cooling (FCC).}
\label{fig:MvT}
\end{figure}

Powder X-ray diffraction (XRD) data were collected for both samples.  Rietveld refinement of the data (see Table~\ref{tab:xrd}) confirmed that both samples had crystallized into the expected orthorhombic structure, with space group \emph{Pnma} and lattice parameters in good agreement with those previously reported.~\cite{Shelton1980}

\begin{table}
\caption{Lattice parameters determined from the Rietveld refinements to the powder-XRD data.}
\centering
\label{tab:xrd}
\begin{ruledtabular}
\begin{tabular}{l c c}
 & \multirow{2}{*}{LuRuB$_2$} & \multirow{2}{*}{YRuB$_2$} \\
\\
\hline
\\
Structure & Orthorhombic & Orthorhombic \\
Space Group & \emph{Pnma} & \emph{Pnma} \\
a (\AA) & \num{5.8075\pm0.0009} & \num{5.9071\pm0.0004} \\
b (\AA) & \num{5.2323\pm0.0007} & \num{5.2971\pm0.0003} \\
c (\AA) & \num{6.2790\pm0.0009} & \num{6.3535\pm0.0004} \\

\end{tabular}
\end{ruledtabular}
\end{table}

The superconducting transition temperature, \tc, for each sample was checked via dc magnetic susceptiility measurements using a \SI{5}{\tesla} Quantum Design Magnetic Property Measurement System.  The temperature dependence of the magnetic susceptibility in an applied field of \SI{1}{\milli\tesla} is displayed in Fig.~\ref{fig:MvT}.  The observed transition temperatures for the Lu and Y compounds are approximately \SI{9.8}{\kelvin} and \SI{7.8}{\kelvin}, in agreement with previous reports.~\cite{Ku1980,Lee1987}  After correcting for demagnetization, a full superconducting volume fraction is found in both samples.  The Meissner fraction, $\chi_\mathrm{FCC}/\chi_\mathrm{ZFCW}$, in the Y compound is 11 times larger than in the Lu compound, indicating that flux pinning is much weaker in YRuB$_2$.  The dc susceptibility data highlights no irregularites or anomalies that may be due to impurities in the sample ordering magnetically or become superconducting.

\subsection{Muon Spectroscopy}

Muon-spin relaxation measurements in zero-field (ZF) and muon-spin rotation experiments in transverse-field (TF) were carried out on the MuSR spectrometer at the ISIS pulsed neutron and muon source, based at the Rutherford Appleton Laboratory in the U.K.  The ISIS synchrotron produces pulses of protons at a frequency of \SI{50}{\hertz}, where \num{4} out of \num{5} pulses pass through the graphite muon production target.  The muons produced in this fashion are \SI{100}{\percent} spin polarized, and after filtering to a momentum of \SI{26}{\mega\electronvolt}/c, are delivered to the MuSR spectrometer where they are implanted into the sample.  The muons rapidly thermalize and sit at interstitial positions in the crystal lattice.  Positive muons decay after an average lifetime of \SI{2.2}{\micro\second} into a positron and two neutrinos, where the positron is emitted preferentially in the direction of the muon spin vector.  The decay positrons are detected and time-stamped in the 64 scintillation detectors, which are arranged in circular arrays positioned before, $F$, or after, $B$, the sample for longitudinal (relaxation) experiments.  The asymmetry $A$ of the \musr\ time spectrum is then calculated by taking the difference of the counts in the $F$ and $B$ detector arrays, weighted by the total number of counts: $A(t) = (F(t)-\alpha B(t))/(F(t)+\alpha B(t))$.  Here $\alpha$ is a calibration constant which represents a relative counting efficiency between the $F$ and $B$ detectors.  The asymmetry function allows one to measure the time evolution of the muon spin polarization, and thus the local magnetic environment experienced by the muon ensemble can be determined.

In a TF experiment, a magnetic field is applied perpendicular to the initial muon spin polarization direction.  In this configuration, the signals from the 64 detectors are normalized and subsequently mapped into two orthogonal components, which are then analysed simultaneously.~\cite{Rainford1999}

Powdered samples were mounted on silver sample plates using GE varnish.  Silver is used as in ZF it produces a time-independent background, whilst in TF it contributes a non-decaying oscillation; both cases are easy to account for during data analysis.  Both samples were mounted in a $^3$He sorption cryostat with a temperature range of \num{0.3} to \SI{50}{\kelvin}.  For the ZF measurements, samples were cooled in zero applied field, and data points were collected in increments upon warming.  Stray fields at the sample position are actively cancelled to within \SI{1}{\micro\tesla} by a flux gate magnetometer and an active compensation system controlling three pairs of correction coils.  The TF experiments were conducted in a field of \SI{30}{\milli\tesla}.  The samples were field cooled to base temperature in order to promote the formation of a well-ordered, flux line lattice, and data points collected upon incremental warming.

\section{Results \& Discussion}

\subsection{Zero \& longitudinal-field muon-spin relaxation}

\begin{figure}[t]
\centering
\includegraphics[width=1.0\columnwidth]{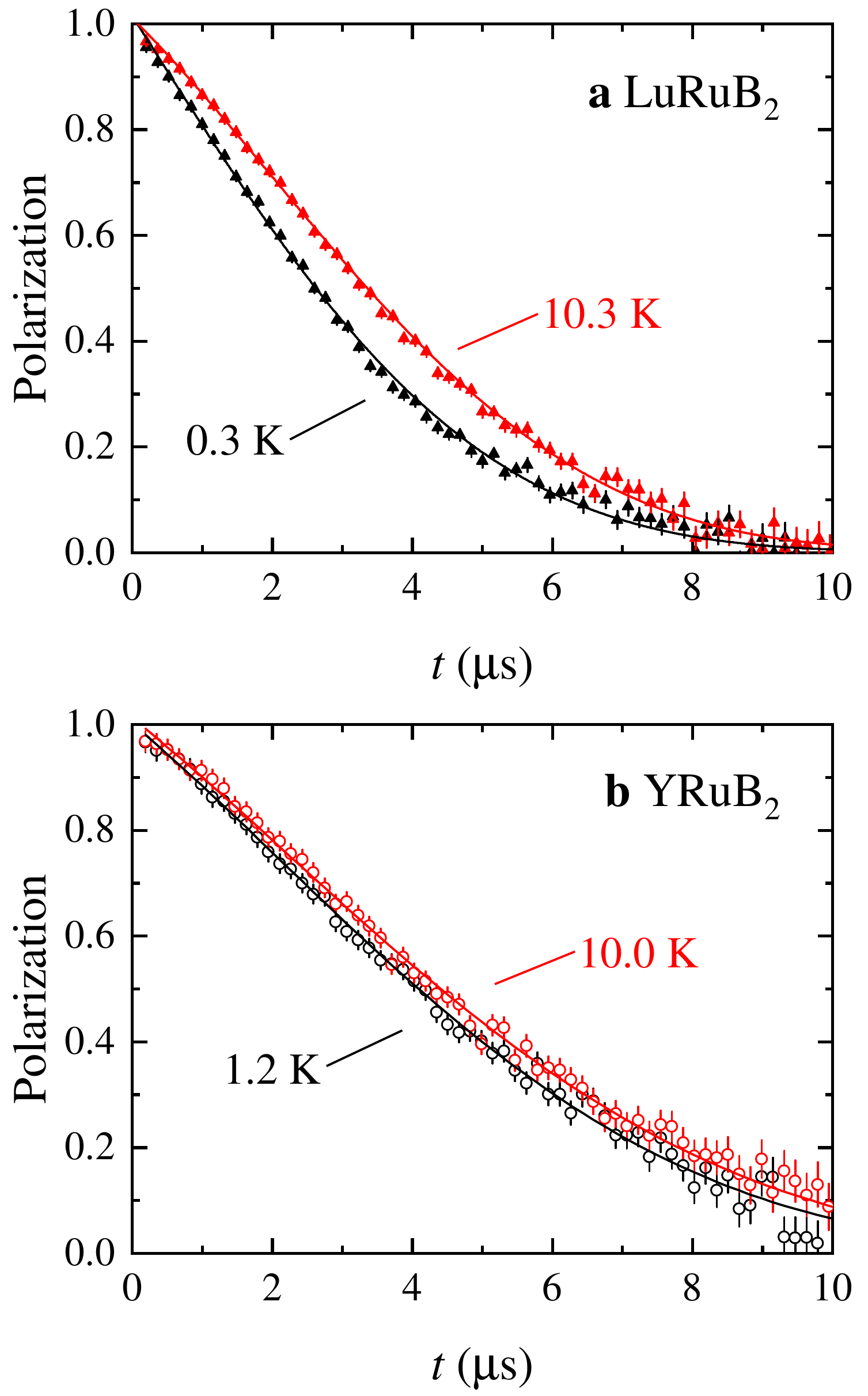}
\caption{(Color online) Time evolution of the spin polarization of muons implanted under zero-field conditions in (a) LuRuB$_2$ and (b) YRuB$_2$ at temperatures above and below \tc.  The time independent background due to muons stopping in silver has been subtracted, and the data normalized to the initial asymmetry - the muons are \SI{100}{\percent} spin-polarized at $t = $~\SI{0}{\second}.  The solid lines are the results of fitting the data to Eq.~\eqref{eq:AS}}
\label{fig:ZFspectra}
\end{figure}

Results from the ZF-\musr\ relaxation experiments are presented first.  Figure \ref{fig:ZFspectra} shows the time evolution of the muon-spin polarization in both samples collected above and below \tc.  There is a clear change in the relaxation behaviour on either side of the transition in both compounds, although the difference is much subtler in the Y compound.  There is no evidence for an oscillatory component, which indicates that there is no coherent field associated with magnetic ordering.  In the absence of atomic moments, the depolarization of the muon ensemble is due to the presence of static, randomly oriented nuclear moments.  This behaviour is modeled by the {Gaussian} Kubo-Toyabe equation~\cite{Hayano1979}
\begin{equation}
\label{eq:KT}
	G_\mathrm{KT}(t) = \frac{1}{3} + \frac{2}{3}(1-\sigma_\mathrm{ZF}^2 t^2)\exp \left(-\frac{\sigma_\mathrm{ZF}^2 t^2}{2} \right),
\end{equation}
where $\sigma_\mathrm{ZF}$ measures the width of the nuclear dipolar field experienced by the muons.  The spectra are well described by the function
\begin{equation}
\label{eq:AS}
	G_z(t) = G_\mathrm{KT}(t) \exp(-\Lambda t),
\end{equation}
where $\Lambda$ measures the electronic relaxation rate, and is usually attributed to `fast-fluctuation' effects that occur on a timescale much shorter than the muon lifetime.

\begin{figure}[t]
\centering
\includegraphics[width=1.0\columnwidth]{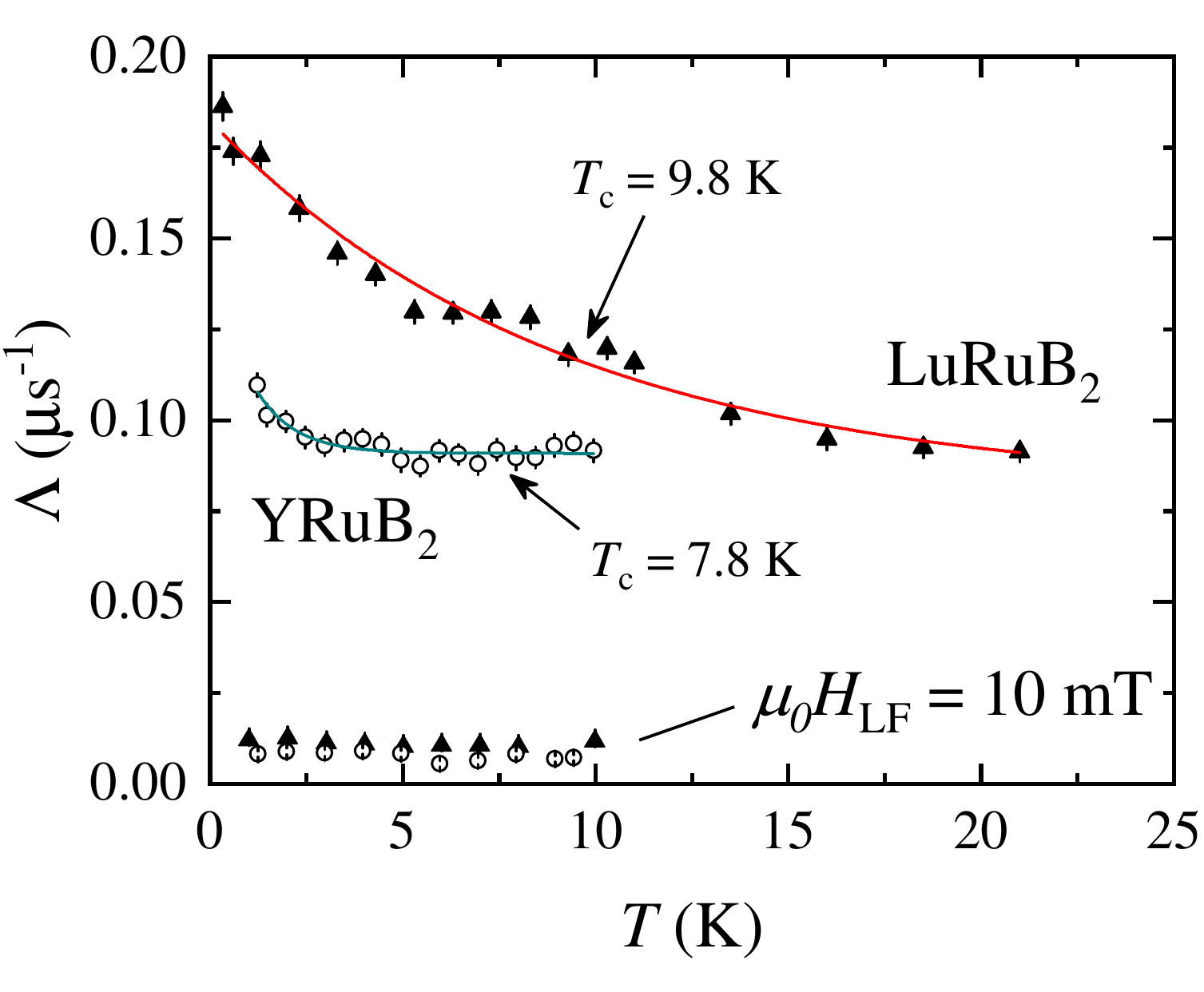}
\caption{(Color online) Temperature dependence of the electronic relaxation rate in LuRuB$_2$ (triangles) and YRuB$_2$ (circles), collected in ZF and in an applied longitudinal field of \SI{10}{\milli\tesla}.  The solid lines are guides to the eye, indicating the exponential decay of $\Lambda$ in ZF as $T$ is increased.}
\label{fig:Lambda}
\end{figure}

The nuclear term $\sigma_\mathrm{ZF}$ is found to remain temperature independent in both compounds.  As the temperature is increased from base, there is an exponential decrease in $\Lambda$ in both materials (see Fig.~\ref{fig:Lambda}).  This is reminiscent of the `critical slowing down' behaviour of spin-fluctuations in the vicinity of phase transitions to magnetically ordered states.~\cite{Dunsiger1996}  In both materials a small longitudinal field of \SI{10}{\milli\tesla} is sufficient to completely decouple the Gaussian component of the relaxation.  Furthermore, the electronic component is almost completely suppressed from the ZF values, implying that the fluctuations reponsible for this relaxation channel are in-fact static or quasistatic with respect to the muon lifetime.  There is no clear anomaly at \tc\ in either material, indicating that the process responsible for these fluctuations is independent of the superconductivity.  Although tempting, we conclude that we do not see any evidence for broken time-reversal symmetry.

\subsection{Transverse-field muon-spin rotation}

\begin{figure}[t]
\centering
\includegraphics[width=1.0\columnwidth]{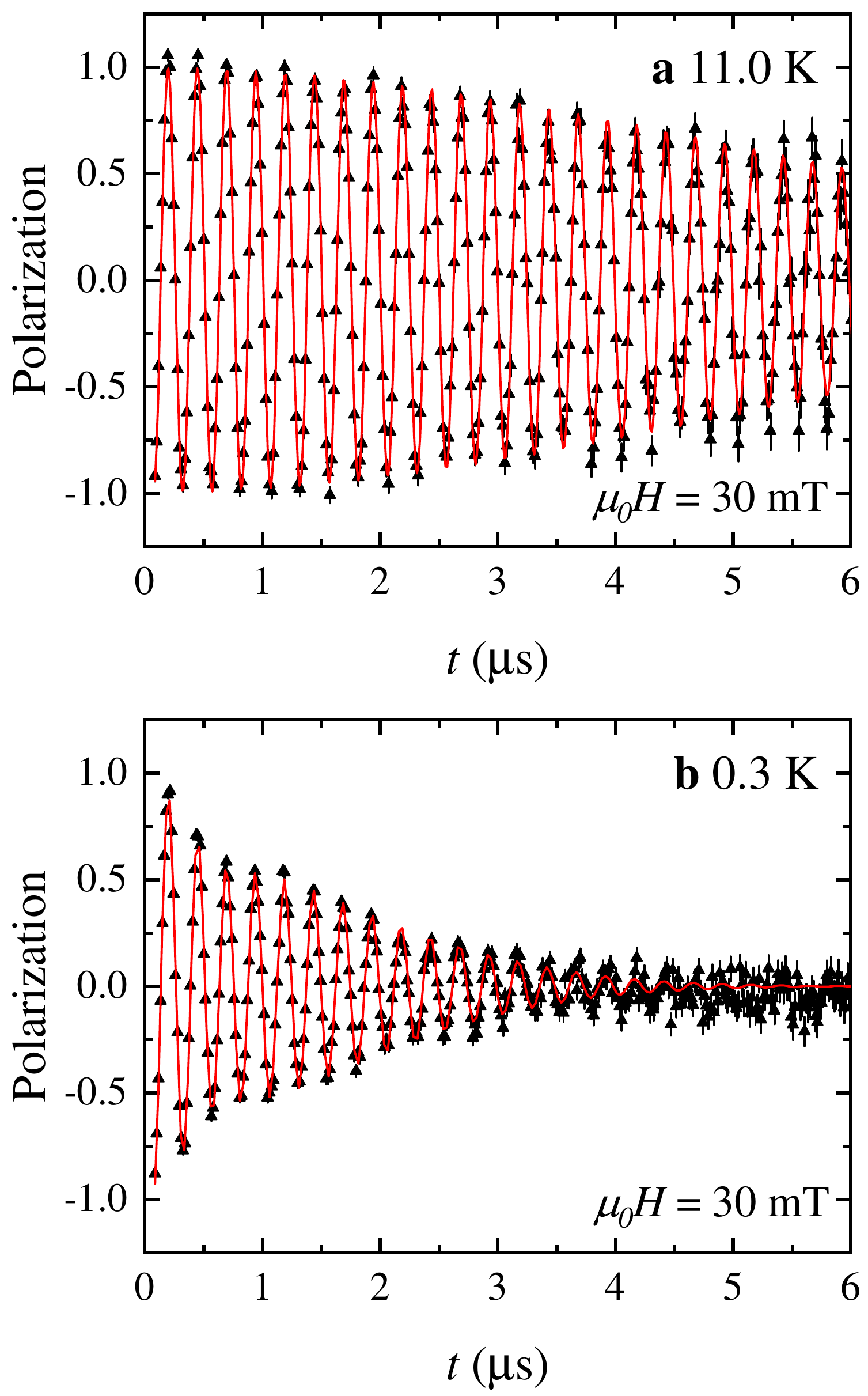}
\caption{(Color online) Representative TF-\musr\ polarization signals collected (a) above and (b) below \tc\ in LuRuB$_2$ under an applied field of \SI{30}{\milli\tesla}.  A non-decaying background oscillation due to muons stopping in the silver has been subtracted, and the data normalised to the initial asymmetry.  The solid lines are fits using Eq.~\eqref{eq:TFfit}.}
\label{fig:TFspectra}
\end{figure}

In order to characterize the flux-line lattice, TF-\musr\ was performed in a field of \SI{30}{\milli\tesla} in both materials.  A selection of typical polarization spectra collected above and below \tc\ is displayed in Fig.~\ref{fig:TFspectra}.  The enhanced depolarization rate below \tc\ is due to the field distribution $P(B)$ formed by the flux line lattice in the mixed state of the superconductor.  Measuring the second moment $\langle \Delta B^2 \rangle$ of this field distribution allows the magnetic penetration depth, $\lambda$, to be calculated to a high degree of accuracy.  In order to determine $\langle \Delta B^2 \rangle$, the TF spectra are modelled as a sum of $n$ sinusoidal oscillations, each within a Gaussian relaxation envelope:
\begin{equation}
	\label{eq:TFfit}
	G_x(t) = \sum_{i=1}^{n}A_i \exp\left(-\frac{\sigma_i^2 t^2}{2}\right)\cos(\gamma_\mu B_i t + \phi),
\end{equation}
where $A_i$ is the initial asymmetry, $\sigma_i$ is the Gaussian relaxation rate, and $B_i$ is the first moment of the $i$'th component in the field distribution.  There is a phase offset $\phi$, which is shared by each oscillating component, and $\gamma_\mu / 2\pi = $~\SI{135.5}{\mega\hertz\per\tesla} defines the muon gyromagnetic ratio.  The number of components required is generally in the range $1 \le n\le 5$, with the requirement determined by the superconducting characteristics of the material.  Strongly type-II superconductors with large penetration depths are often modelled well by a single oscillation, whereas low-$\kappa$ materials, in which the coherence length plays a more important role in the structure of $P(B)$, may require up to \num{5} separate oscillating components.~\cite{Khasanov2008}  Treating the data in this way is equivalent to modelling the internal field distribution in the superconductor $P(B)$ as a sum of $n$ individual Gaussians,~\cite{Maisuradze2010}
\begin{equation}
	\label{eq:pb}
	P(B) = \gamma_\mu \sum_{i=1}^{n} \frac{A_i}{\sigma_i}\exp\left(-\frac{\gamma_\mu^2(B-B_i)^2}{2\sigma_i^2}\right).
\end{equation}
The second moment of this field distribution is thus
\begin{equation}
	\label{eq:db2}
	\langle\Delta B^2 \rangle = \frac{\sigma_\mathrm{eff}^2}{\gamma_\mu^2} = \sum_{i=1}^{n} \frac{A_i}{A_\mathrm{tot}}\left[\frac{\sigma_i^2}{\gamma_\mu^2} + (B_i - \langle B \rangle )^2 \right],
\end{equation}
where $A_\mathrm{tot} = \sum_{i=1}^{n} A_i$ and $\langle B \rangle = A_\mathrm{tot}^{-1} \sum_{i=1}^{n} A_i B_i$ is the first moment of $P(B)$.  Finally, the extra broadening from the nuclear moments $\sigma_\mathrm{N}$ must be subtracted in quadrature from the total effective depolarization rate $\sigma_\mathrm{eff}$ to yield the contribution of the flux-line lattice $\sigma_\mathrm{FLL}^2 = \sigma_\mathrm{eff}^2 - \sigma_\mathrm{N}^2$.  $\sigma_\mathrm{N}$ is assumed to be temperature independent, and is determined by measurements made in the normal state just above \tc.  Two oscillating components were required to adequately describe the LuRuB$_2$ spectra, whereas three were required for the YRuB$_2$ - a non-decaying background oscillation due to muons stopping in the silver sample holder has been subtracted from the spectra presented in Fig.~\ref{fig:TFspectra}.  Above \tc\ a single oscillation suffices in both materials to describe the depolarization.  



\begin{figure}[t]
\centering
\includegraphics[width=1.0\columnwidth]{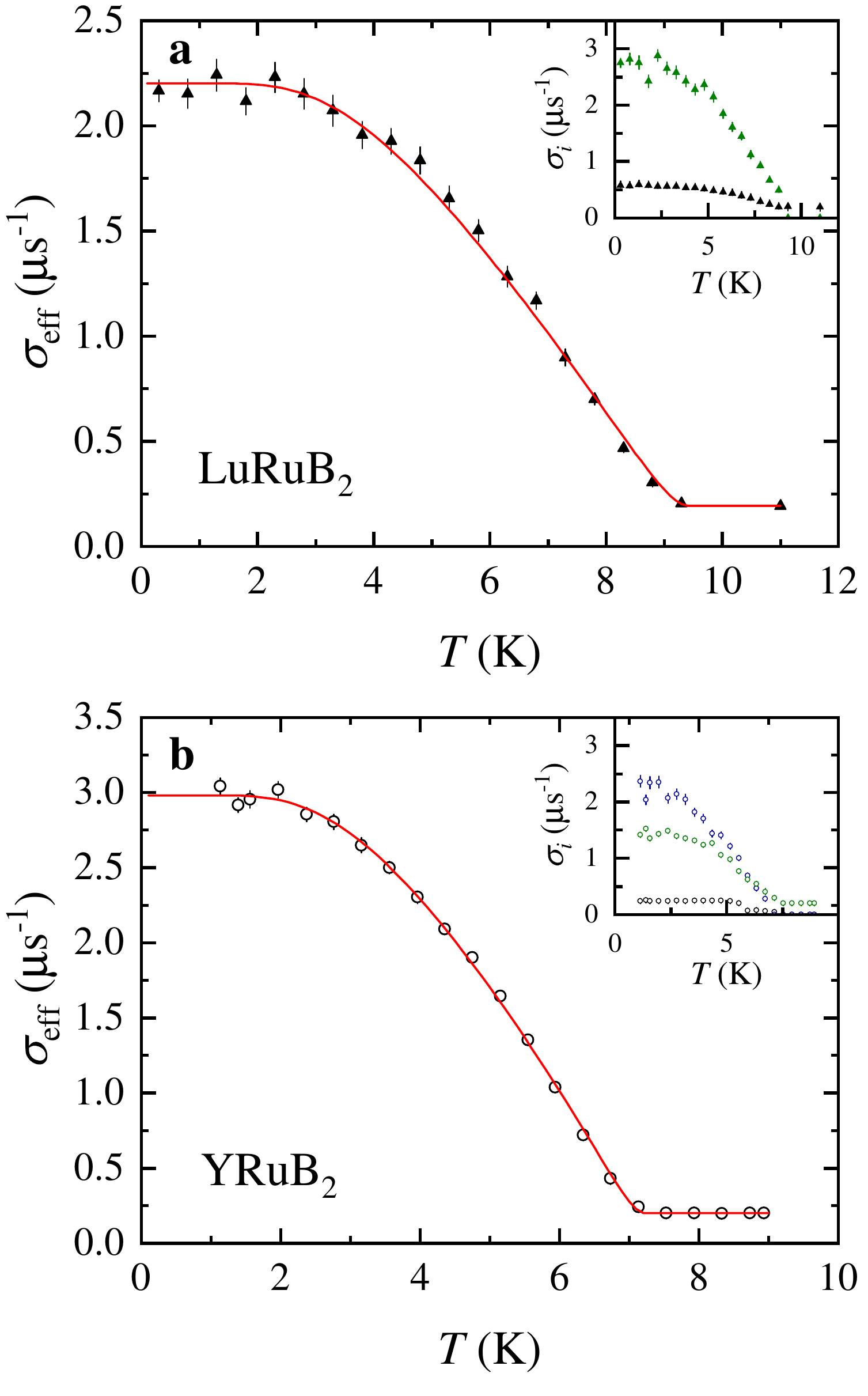}
\caption{(Color online) TF-\musr\ effective depolarization rates in (a) LuRuB$_2$ and (b) YRuB$_2$, calculated from the $\sigma_i$ (insets) as described in the text.  The solid line is a fit to Eq.~\ref{eq:pendepclean}, which is valid as there is a simple numerical coefficient relating $\sigma_\mathrm{eff}$ and $\lambda^{-2}$.}
\label{fig:depolarization}
\end{figure}

The temperature dependences of $\sigma_\mathrm{eff}$ in both compounds are presented in Fig.~\ref{fig:depolarization}.  In superconductors with large critical fields and hexagonal flux line lattices, there exists a simple relationship between the Gaussian depolarization rate $\sigma_\mathrm{FLL}$ and the magnetic penetration depth, as long as the average field is a very small fraction of the upper critical field \ucf.  For both compounds $B/$\ucf$\approx 0.005$ and so we can use the expression~\cite{Brandt2009}
\begin{equation}
	\label{eq:brandt}
	\frac{\sigma_\mathrm{FLL}^2(T)}{\gamma_\mu^2} = 0.00371\frac{\Phi_0^2}{\lambda^4(T)},
\end{equation}
where $\Phi_0$ is the magnetic flux quantum.  The magnetic penetration depths at $T = $~\SI{0}{\kelvin} are thus found to be $\lambda_\mathrm{Lu}(0) = $~\SI{221\pm2}{\nano\metre} and $\lambda_\mathrm{Y}(0) = $~\SI{190\pm1}{\nano\metre} for the LuRuB$_2$ and YRuB$_2$ materials, respectively.

\begin{figure}[t]
\centering
\includegraphics[width=1.0\columnwidth]{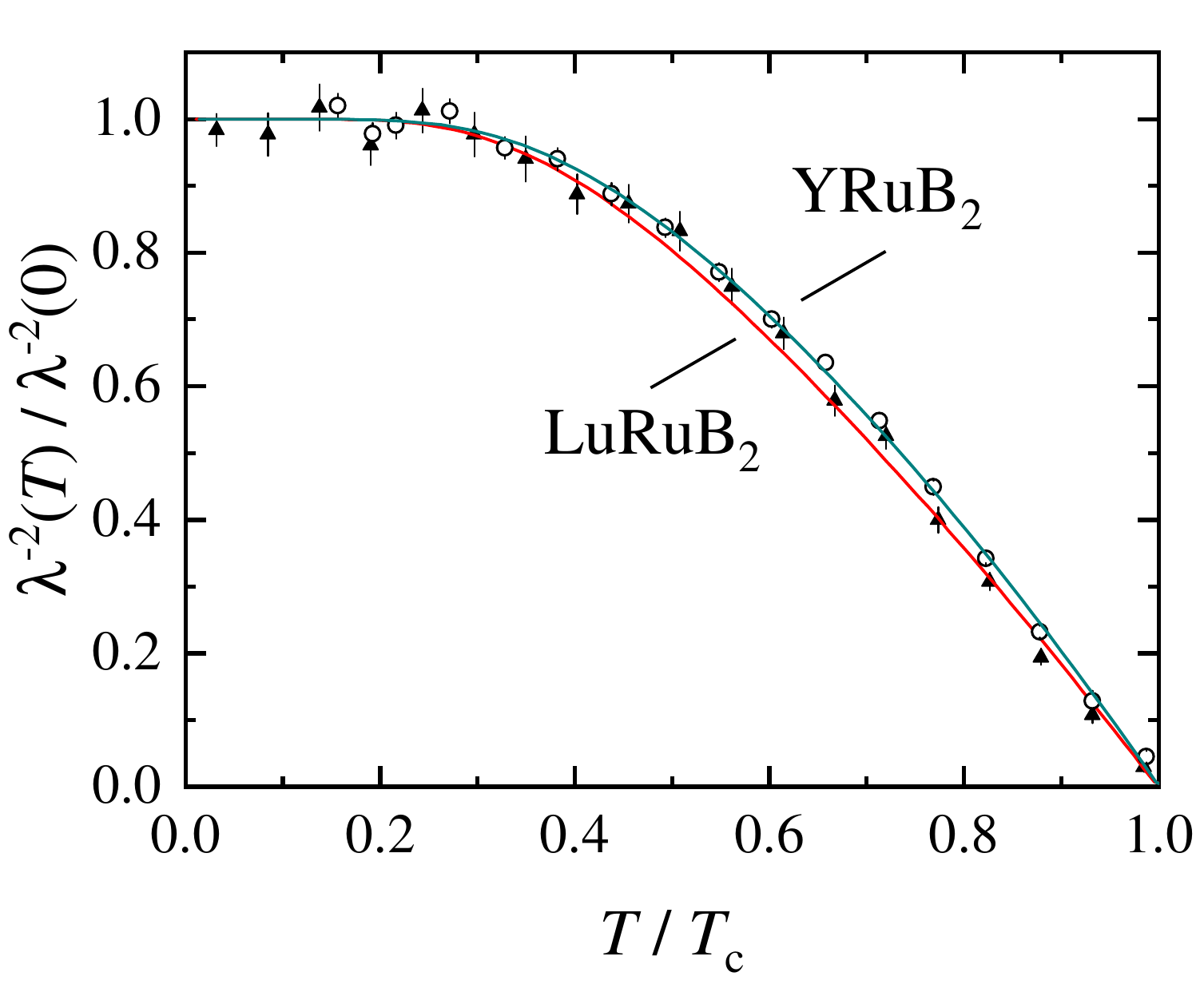}
\caption{(Color online) Temperature dependence of the superfluid density as a function of the reduced temperature $T/T_\mathrm{c}$ for LuRuB$_2$ (triangles) and YRuB$_2$ (circles).  The data overlay each other, reflecting the high degree of similarity in the order parameters of the two materials.  The solid lines are fits using Eq.~\eqref{eq:pendepclean}.}
\label{fig:pendeps}
\end{figure}

Assuming London local electrodynamics, the temperature dependence of $\lambda$ can be calculated for an isotropic \emph{s}-wave superconductor in the clean limit using the following equation:
\begin{equation}
\label{eq:pendepclean}
	\frac{\lambda^{-2}(T)}{\lambda^{-2}(0)} = 1 + 2 \int_{\Delta(T)}^{\infty} \left(\frac{\partial f}{\partial E}\right)\frac{E~\!  \mathrm{d}E}{\sqrt{E^2 - \Delta^2(T)}},
\end{equation}
where $f=[1+\exp(E/k_\mathrm{B} T)]^{-1}$ is the Fermi function and $\Delta(T) = \Delta(0)\tanh\{1.82[1.018(T_c/T - 1)]^{0.51}\}$ is the BCS approximation for the temperature dependence of the energy gap.  The normalized inverse-squared penetration depth, or superfluid density, is displayed in Fig.~\ref{fig:pendeps} for both materials, with fits to the data using this model.  The resultant values for the energy gaps are $\Delta_\mathrm{Lu}(0)=$~\SI{1.36\pm0.03}{\milli\electronvolt} and $\Delta_\mathrm{Y}(0)=$~\SI{1.10\pm0.01}{\milli\electronvolt}.  The BCS theory proposes a universal proportionality between the energy gap and the superconducting transition temperature.  Ths is conventionally encoded in the BCS parameter, $2\Delta(0)/k_\mathrm{B} T_\mathrm{c}$, which has the theoretical value of \num{3.52} in the weak coupling limit.  For the Lu and Y compounds, the BCS parameters are found to be \num{3.3\pm0.2} and \num{3.4\pm0.1}, respectively.  This seems to classify the (Lu/Y)RuB$_2$ ternary borides as conventional, weakly coupled BCS type-II superconductors, in agreement with the NMR results.~\cite{Kishimoto2009}

The magnetic penetration depth is directly related to the electronic properties of the superconducting state by the expression\cite{Tinkham2004}
\begin{equation}
\label{eq:lambda0}
	\lambda(0) = \left[\frac{m^*}{\mu_0 n_\mathrm{s} e^2}\left(1+\frac{\xi_0}{l}\right) \right]^{\frac{1}{2}},
\end{equation}
where $m^*$ is the effective mass of charge carrying electrons, and $n_\mathrm{s}$ is the superconducting charge carrier density. The ratio of BCS coherence length to the mean free path, $\xi_0/l$, encodes the dirty limit correction, which for the Lu and Y compounds has been found to take on the values  \num{3.9} and \num{0.85}, respectively.~\cite{Lee1987}  Equation~\eqref{eq:lambda0} can be coupled with the expression for the Sommerfeld constant $\gamma$, which is also related to the electronic properties of the system:~\cite{Hillier1997}
\begin{equation}
\label{eq:gamma}
	\gamma = \left( \frac{\pi}{3}\right)^{\frac{2}{3}} \frac{k_\mathrm{B}^2 m^* n_\mathrm{e}^{\frac{1}{3}} }{\hbar^2},
\end{equation}
where $n_\mathrm{e}$ is the electronic carrier density and $k_\mathrm{B}$ is Boltzmann's constant.  By assuming that $n_\mathrm{e}$ at \tc\ is equivalent to $n_\mathrm{s}$ as $T \rightarrow $~\SI{0}{\kelvin}, Eqs.~\eqref{eq:lambda0} and \eqref{eq:gamma} can be solved simultaneously to find values for $m^*$ and $n_\mathrm{s}$.  Consequently an effective Fermi temperature can be calculated using the relation $k_\mathrm{B} T_\mathrm{F} = (\hbar^2/2)(3\pi^2n_\mathrm{s})^{2/3}/m^*$.  The results of following this procedure are displayed in Table~\ref{tab:results}.

Uemura \emph{et al.}\ have described a method of classifying superconductors based on the ratio of the critical temperature \tc\ to the effective Fermi temperature $T_\mathrm{F}$, which is found to be $1/414$ and $1/304$ for the Lu and Y compounds, respectively.~\cite{Uemura1988,Uemura1989,Uemura1991}  This places the ternary borides in the vicinity of the `band of unconventionality' described by Uemura.  This is the first indication that the superconductivity in these compounds may not be entirely conventional.  In fact, both compounds find themselves occupying the same region in the Uemura diagram as the borocarbide superconductors, and the rare-earth hexaborides.~\cite{Hillier1997}  High transition temperatures are a common theme in these families of materials, as well as the intriguing interplay between the superconductivity and the complex magnetic order associated with the rare-earth $4f$ electrons.

\begin{figure}[t]
\centering
\includegraphics[width=1.0\columnwidth]{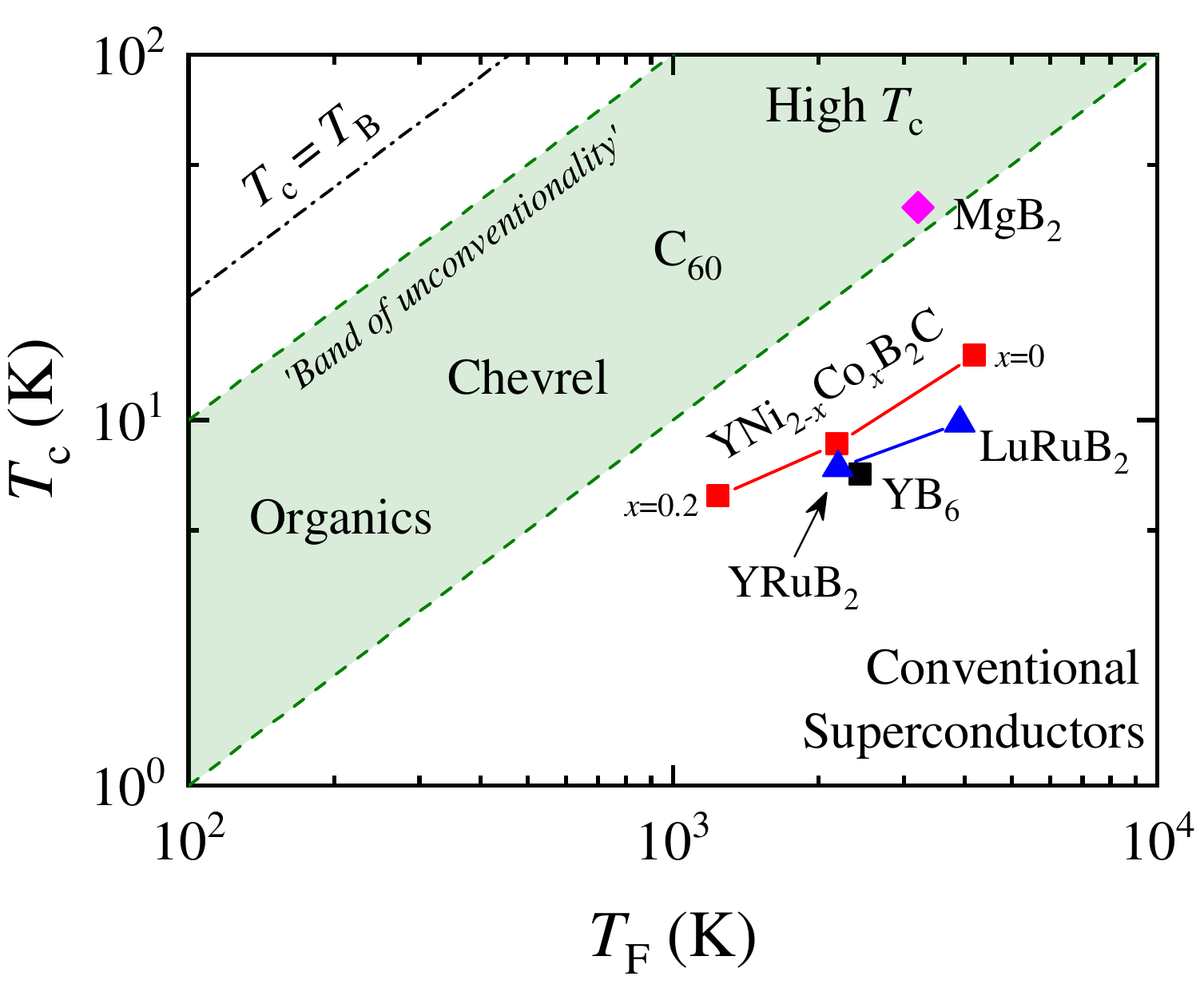}
\caption{(Color online) The results of the \musr\ experiments summarized in the `Uemura plot', which describes a universal scaling between \tc\ and $T_\mathrm{F}$ in different classes of superconductors.  The Lu and Y ternary borides find themselves halfway between the conventional and unconventional regions, in the vicinity of the borocarbide and hexaboride superconductors.}
\label{fig:uemura}
\end{figure}

\begin{table}
\caption{Superconducting properties determined from the TF-\musr\ experimental results.}
\centering
\label{tab:results}
\begin{ruledtabular}
\begin{tabular}{l c c}
 & \multirow{2}{*}{LuRuB$_2$} & \multirow{2}{*}{YRuB$_2$} \\
\\
\hline
\\
$\lambda$ (\si{\nano\meter}) & \num{221\pm2} & \num{190\pm1} \\
$\Delta$(0) (\si{\milli\electronvolt}) & \num{1.36\pm0.03} & \num{1.10\pm0.01} \\
BCS parameter & \num{3.3\pm0.2} & \num{3.4\pm0.1} \\
$m^*/m_\mathrm{e}$ & \num{9.8\pm0.1} & \num{15.0\pm0.1} \\
$n_\mathrm{s}$ ($\times 10^{28}$~\si{\per\meter\cubed}) & \num{2.73\pm0.04} & \num{2.17\pm0.02} \\
\tc$/T_\mathrm{F}$ & $1/(414\pm6)$ & $1/(304\pm3)$ \\

\end{tabular}
\end{ruledtabular}
\end{table}

\section{Conclusions}

In conclusion, TF and ZF-\musr\ measurements have been carried out on the rare-earth ternary borides (Lu/Y)RuB$_2$.  Both superconductors are well described by the conventional BCS theory of superconductivity in the weakly coupled limit, with fully gapped \emph{s}-wave order parameters and preserved time-reversal symmetry in the superconducting state.  The ZF-\musr\ measurements reveal spin fluctuations that exhibit a critical slowing down behaviour as the temperature is decreased, implying that both systems may be close to quantum critical points.  Calculations of the electronic properties of the superconducting state reveal that the rare-earth ternary borides share similarities with the hexaboride and borocarbide superconducting families.

\begin{acknowledgments}
The authors would like to thank Mr T. E. Orton for valuable technical support.  J.~A.~T.~B.\ acknowledges ISIS and the STFC for studentship funding through grant ST/K502418/1.  This work was funded by the EPSRC, $\mathrm{U.K.}$, through grant EP/I007210/1. Some of the equipment used in this research was obtained through the Science City Advanced Materials project: Creating and Characterizing Next Generation Advanced Materials project, with support from Advantage West Midlands and part funded by the European Regional Development Fund.
\end{acknowledgments}

\end{document}